# Integrating uncertainty in deep neural networks for MRI based stroke analysis

**Lisa Herzog**[1,2*] **Elvis Murina**[2] **Oliver Dürr**[3] **Susanne Wegener**[4] **Beate Sick**[1,2*]

[1]University of Zurich (UZH), [2]Zurich University of Applied Sciences (ZHAW), [3]Konstanz University of Applied Sciences, [4]University Hospital Zurich (UHZ)

**Abstract**

At present, the majority of the proposed Deep Learning (DL) methods provide point predictions without quantifying the model's uncertainty. However, a quantification of the reliability of automated image analysis is essential, in particular in medicine when physicians rely on the results for making critical treatment decisions. In this work, we provide an entire framework to diagnose ischemic stroke patients incorporating Bayesian uncertainty into the analysis procedure. We present a Bayesian Convolutional Neural Network (CNN) yielding a probability for a stroke lesion on 2D Magnetic Resonance (MR) images with corresponding uncertainty information about the reliability of the prediction. For patient-level diagnoses, different aggregation methods are proposed and evaluated, which combine the single image-level predictions. Those methods take advantage of the uncertainty in image predictions and report model uncertainty at the patient-level. In a cohort of 511 patients, our Bayesian CNN achieved an accuracy of 95.33% at the image-level representing a significant improvement of 2% over a non-Bayesian counterpart. The best patient aggregation method yielded 95.89% of accuracy. Integrating uncertainty information about image predictions in aggregation models resulted in higher uncertainty measures to false patient classifications, which enabled to filter critical patient diagnoses that are supposed to be closer examined by a medical doctor. We therefore recommend using Bayesian approaches not only for improved image-level prediction and uncertainty estimation but also for the detection of uncertain aggregations at the patient-level.

## 1. Introduction

Ischemic stroke elicits neurological deficits with sudden onset caused by reduced blood flow and oxygen supply to the brain due to occlusion of cerebral blood vessels (Purves, et al., 2004). As an estimate, 1.9 million neurons are lost with every passing minute in patients with large vessel occlusion stroke (Saver, 2015). Accordingly, acute ischemic stroke treatment aims at recanalization of the occluded vessel and restoration of blood flow as fast as possible to prevent disability or death. Besides clinical and patient data, medical experts rely on multi-modal Computed Tomography (CT) and Magnetic Resonance (MR) imaging for diagnoses and treatment decisions (Baird & Warach, 1998; Chilla, et al., 2015). To both support physicians in their diagnosis and accelerate treatment decision-making, methods for automatic image analysis are increasingly integrated into acute stroke care (Feng, et al., 2018; Pinto, et al., 2018). Deep Convolutional Neural Networks (CNNs) are state-of-the-art to recognize pathological features such as ischemic stroke lesions on brain images (Bernal, et al., 2019; Havaei, et al., 2017). In a hierarchy of convolutional operations, increasingly complex representations of data are learnt avoiding the challenging task of feature engineering performed in traditional imaging strategies (Goodfellow, et al., 2016). The data acquisition process of CT and MR images yields an ordered stack of 2D images. Thus, a common choice is to use a 2D CNN to analyze the 2D images, followed by an aggregation method combining resulting image features and predictions to a patient-level outcome (Islam & Zhang, 2017; Patel, et al., 2019; Setio, et al., 2016). In addition, 3D CNNs have shown promising results when analyzing CT and MR images (Kamnitsas, et al., 2015; Lisowska, et al., 2017). However, choosing a 2D over a 3D CNN architecture has several advantages: the common DL frameworks are optimized for multi-channel 2D images, are computationally less expensive, need fewer memory and require smaller sample sizes, which is essential in the presence of medical data. In addition, the performance between 2D and 3D CNNs is oftentimes similar (Lai, 2015; Payan & Montana, 2015).
In recent years, CNNs achieved outstanding performances on medical image analysis tasks comparable to those of medical experts (Kamnitsas, et al., 2016; Rajpurkar, et al., 2017). Yet, most of the proposed CNN architectures lack information about reliability of predictions limiting their applicability in medical care. Moreover, CNNs tend to predict too high probabilities for ambiguous or unknown cases, which are commonly seen in medicine. Reliable uncertainty estimates enable to filter such cases (Dürr, et al., 2018; Leibig, et al., 2017; Nair, et al., 2020), which may subsequently be returned to physicians for additional inspection. In addition, image-level uncertainties can be used to improve the analysis procedure when combining image predictions to a patient diagnosis. As a physician would do, the focus could shift towards images, about which the CNN is most confident.
Bayesian approaches are the mainstay to quantify uncertainty in machine learning. The goal is to estimate the posterior distribution, which represents all plausible predictions and gives insights into the uncertainty. Yet, due to the high-dimensional parameter space of modern CNNs, an analytical computation of the posterior is unfeasible. A recent approach, however, enables to obtain an approximating solution using existing dropout-variants of neural networks; Monte Carlo (MC) dropout (Gal & Ghahramani, 2016). In a standard, non-Bayesian setting, units of neural network layers are randomly inactivated during training. At test time, the entire network is used for outcome prediction (Srivastava, et al., 2014). However, keeping dropout turned on also at test time has been shown to be equivalent to drawing samples of the approximate posterior, which enables to do Bayesian inference (Gal & Ghahramani, 2016). The simplicity and adaptability of this Bayesian approach to all dropout-variants of neural networks, qualifies the technique as an eligible candidate to acquire uncertainty information from CNNs also in the field of medical image analysis (Leibig, et al., 2017; Nair, et al., 2020; Ozdemir, et al., 2017).

In this work, we provide a whole framework to diagnose ischemic stroke patients based on MRI data, which integrates Bayesian uncertainty into the analysis procedure. It mainly consists of two components: a Bayesian CNN for ischemic stroke lesion identification on 2D MR images and advanced aggregation methods to combine the image predictions to patient diagnoses. First, we show that a Bayesian model (1Ch-MC) yields an increased prediction performance compared to its non-Bayesian counterpart (1Ch-NoMC) while simultaneously providing uncertainty estimates. Then we adapt the Bayesian CNN, to consider the special 3D data structure of MR images (3Ch-MC). In the simplest version of a baseline aggregation model, image-level predictions are combined to a patient-level diagnosis, taking the largest stroke probability over all images of a patient. In a next step, we highlight the improvement in prediction performance when using neural network-based aggregation methods. Those methods consider the uncertainty in the image predictions and provide a measure of uncertainty at the patient-level.

## 1.1 Related work

Several approaches integrated MC dropout in CNNs for uncertainty estimation and performance improvement. As way of examples, (Kendall, et al., 2017) proposed one of the first MC dropout networks for improved semantic segmentation tasks. (Zhao, et al., 2018) applied the same network to brain segmentation problems. (Kwon, et al., 2020) used MC dropout to analyze ischemic stroke lesion segmentations and (Leibig, et al., 2017) assessed the approach in diabetic retinopathy. (Nair, et al., 2020) applied it to MS lesion detection and segmentation while (Ozdemir, et al., 2017) integrated MC dropout to improve pulmonary nodule detection.
In the majority of the approaches, uncertainty was quantified though the variance of the predictive distribution resulting from the repeating MC dropout runs (Kendall, et al., 2017; Leibig, et al., 2017; Nair, et al., 2020; Ozdemir, et al., 2017; Tousignant, et al., 2019; Zhao, et al., 2018). (Nair, et al., 2020) additionally evaluated the predictive entropy and mutual information as uncertainty measures. In the original work of (Gal, 2017), uncertainty in classification tasks was described using the predictive entropy, mutual information and variation ratio. More recently, uncertainty was further decomposed into epistemic and aleatoric uncertainty. Epistemic uncertainty represents the noise in the model parameters, i.e. uncertainty due to lack of understanding and availability of training data. Aleatoric uncertainty models the noise inherent in the input data itself. To model both types of uncertainty jointly, (Kendall & Gal, 2017) presented a unified Bayesian framework integrating aleatoric uncertainty as learned loss attenuation. (Kwon, et al., 2020) extended on this idea in a classification setting and decomposed the variance of the variational predictive distribution. (Ryu, et al., 2019) applied this approach to molecular property prediction. Furthermore, test-time augmentation was proposed to be used for aleatoric uncertainty estimation (Ayhan & Berens, 2018; Wang, et al., 2019).

Table 1. **Summary of the patient cohort.** Frequencies of ischemic stroke and transient ischemic attack (TIA) patients as well as corresponding images with (stroke) and without visible stroke lesions (no-stroke).

| | | **Patients** | | |
|---|---|---|---|---|
| | | Stroke (n=355) | TIA (n=156) | |
| **Images** | Stroke | 3120 | - | 3120 |
| | No-stroke | 7412 | 4656 | 12068 |
| | | 10532 | 4656 | 15188 |

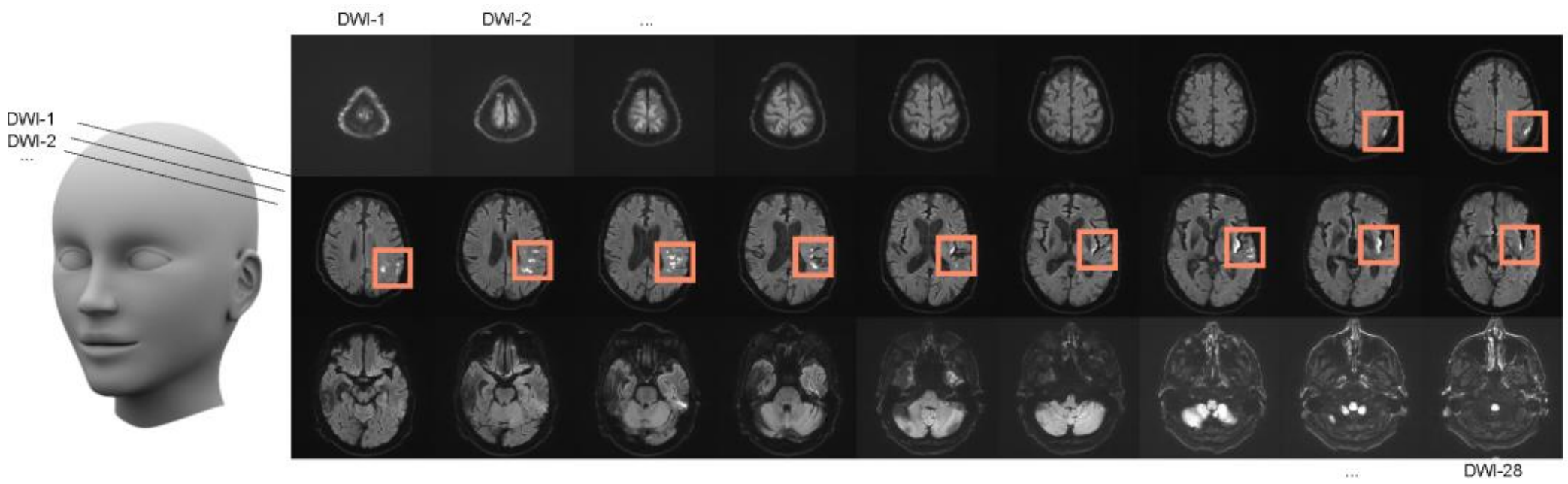

Figure 1. **DW-MRIs of an ischemic stroke patient.** Each patient is represented by a collection of 2D MRIs that show the brain pathology in a 3D manner. If present, the ischemic event appears as white spot, typically on a sequence of slices (rectangles).

In many MC dropout applications, a correlation between high uncertainty estimates and erroneous predictions was found as in phenotype (Dürr, et al., 2018) and diabetic retinopathy classification (Leibig, et al., 2017), pulmonary nodule (Ozdemir, et al., 2017) and multiple sclerosis detection (Nair, et al., 2020) as well as in brain segmentation tasks (Wang, et al., 2019; Zhao, et al., 2018). Yet, there is only little literature making use of that correlation to filter uncertain cases while aiming to improve eventual outcome prediction. (Ozdemir, et al., 2017) implemented a 2D CNN yielding voxel-wise segmentation uncertainty maps. Those maps were combined with the original image to improve pulmonary nodule detection. (Roy, et al., 2018) proposed uncertainty estimates for structure-wise uncertainty based on voxel-wise uncertainties from brain segmentation maps. The uncertainties were then integrated as weights in linear regression models for improved group analysis. (Nair, et al., 2020) applied a 3D CNN for MS lesion segmentation providing voxel-wise predictions and uncertainties. Summing up the voxel-wise log uncertainties yielded a final uncertainty measure for the MS lesion.

Yet, to the best of our knowledge, there is no approach integrating Bayesian uncertainty into the analysis procedure of ischemic stroke patient diagnosis. Moreover, we found no existing aggregation method, which could directly be transferred to the problem of ischemic stroke patient classification. This is because the method has to (i) account for the number of 2D images, which varies substantially between patients, (ii) consider that most of the images are non-discriminative, as even among stroke patients, the majority of images has no lesion, and (iii) provide uncertainty information at the patient level.

## 2. Material and Methods

### 2.1 Data

We retrospectively collected MR images of 511 patients with ischemic stroke or transient ischemic attack (TIA) admitted to the Department of Neurology of the University Hospital Zurich, Switzerland. TIA patients were defined as having similar neurological deficits as stroke patients, which, however, vanish within 24 h and release no visible lesion on the MRI. The MRI modality used within this study was Diffusion Weighted Magnetic Resonance Imaging (DW-MRI) being the primary assessment criteria to distinguish stroke from TIA in acute care (Chilla, et al., 2015). The orientation of the brain on the MR images was the same among patients but sizes and spatial locations of lesions varied substantially emphasizing the incorporation of mild to severe ischemic vessel occlusions. Ethical approval for the study was obtained from the cantonal Ethics Committee Zurich, reassuring the performance of the study according to ethical guidelines (KEK-ZH-No. 2014-0304).

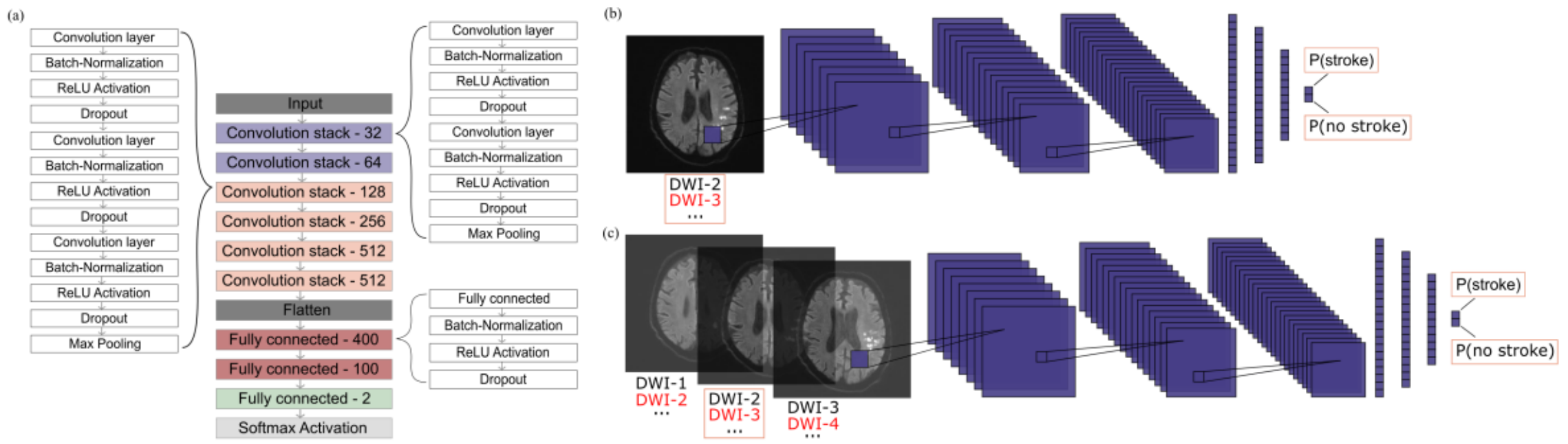

Figure 2. **CNN architectures for image-level predictions.** (a) Each CNN consists of a sequence of layers stacked on each other to differentiate stroke from no-stroke images. (b) As baseline benchmark, we fed single, 1D images into the network and analysed it once with standard dropout (1Ch-NoMC) and once with MC Dropout (1Ch-MC). (c) The 3D data structure was considered, when we generated 3-channel input images consisting of the predictable image (second channel) and the images above (first channel) and below (third channel). This model is referred to as 3Ch-MC.

An experienced neurologist checked all images of stroke and TIA patients on the presence of lesions to define ground-truth labels for each image, namely "stroke" and "no-stroke". Although this is a rather simple problem, in case of doubt, the DW-MRIs were compared to their counterparts, the Apparent Diffusion Coefficient (ADC) maps, on which a true lesion appears hypo- instead of hyper-intense. Furthermore, radiology reports were available. Prior to the image labelling process by the neurologist, a neuroradiologist and a physician in training screened the images and diagnosed the patients as "stroke" and "TIA". If the DW-MRI, the ADC map and the radiology report did not provide enough information, the image was excluded (55 images in total). Consequently, we expect the label uncertainty in our data set to be minimal.

On average, there were 30 images per patient, ranging between a minimum of 21 and a maximum of 46. The average number of images representing a stroke lesion was 12.5. In total, 3120 stroke lesions on 10532 images were related to 355 ischemic stroke patients; the 156 TIA patients were represented by 4656 no-stroke images (see Table 1). Notice, even for patients with ischemic stroke, most of the images had no visible lesion, which led to an imbalance among stroke and no-stroke images in the cohort (see Figure 1). Dimension of the 2D DW-MRIs was mostly 192x192x1 pixels. Images with higher or lower pixel counts in spatial structure were rescaled to 192x192 pixel greyscales. Black images containing no information were removed from the data set. Each image was normalized to have zero mean and unit variance. No further pre-processing was performed.

## 2.2 CNN architectures for image-level predictions

Our CNN architecture to differentiate between stroke and no-stroke images (see Figure 2 (a)) was inspired by the VGG (Simonyan & Zisserman, 2014), which has been shown to perform well in many different applications (Kendall, et al., 2017; Leibig, et al., 2017; Tousignant, et al., 2019). The convolutional part consists of six blocks with two respectively three convolutional layers, which have been added iteratively until the validation accuracy leveled off. Each block completes with a max-pooling layer (window size 2x2 pixels, stride width 2). The number of filters was doubled with every block except the last one while the filter size was fixed to 3x3 pixels. The fully connected part comprises a 400-, a 100- and a two-unit fully connected layer with softmax activation attached. After each convolutional and fully connected layer, batch-normalization layers were included (Ioffe & Szegedy, 2015). Starting with the second convolutional layer, dropout layers (with dropout probability 0.3) were inserted before each convolutional and fully connected layer (Srivastava, et al., 2014). In addition, we applied dropout on the input but found this to reduce the performance of our CNN significantly. As activation function, ReLU non-linearity was used (Nair & Hinton, 2010).

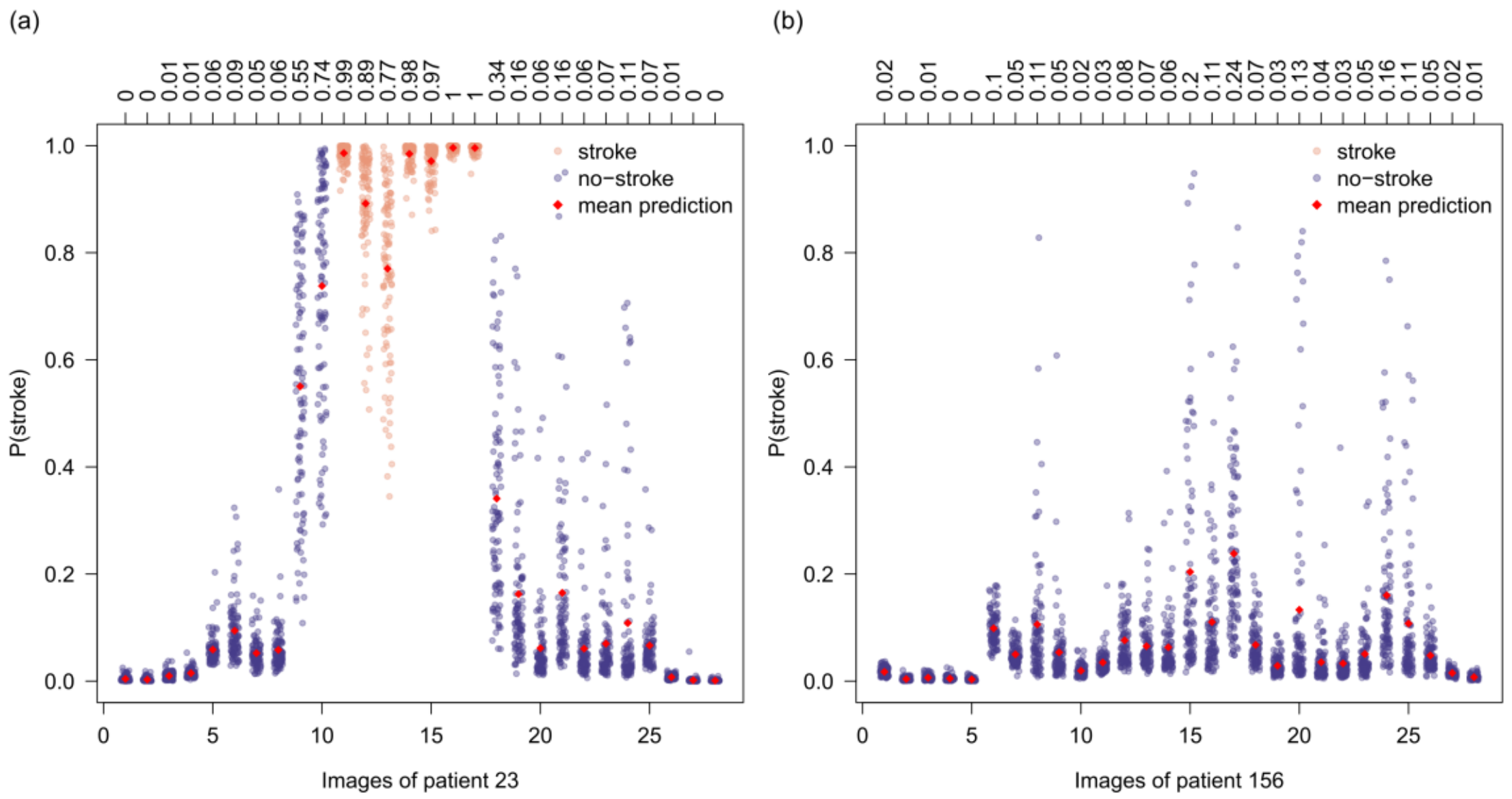

Figure 3. **Image-level predictions for a stroke and a TIA patient resulting from MC Dropout.** The figure summarizes the predictions for stroke patient 23 and TIA patient 156 obtained with the 3Ch-MC. Applying MC Dropout yields, for each image of each patient, a whole distribution of probabilities that the image represents a stroke lesion (P(stroke)). The probabilities result from 500 MC dropout runs. A large spread of the distribution indicates a high prediction uncertainty. The colours highlight the ground-truth labels of the images, namely stroke and no-stroke. The mean predicted probability is marked with a dark red diamond and reported on the top axis.

3D information was integrated when we generated 3-channel (3Ch) images consisting of the image of interest and two adjacent images (see Figure 2 (c)). The predictable image was inserted in the second channel along the depth, images above and below in the first and third depth-channel, respectively. As stroke lesions typically appear on a sequence of images within the MRI volume (see Figure 1), neighboring information is supposed to support CNN classification. In the following, the model is referred to as 3Ch-MC. As benchmark for the relative performance of our model, a CNN featuring the same architecture but expecting single, one-dimensional images without neighboring information was considered (1Ch-MC) (see Figure 2 (b)). The impact of using a Bayesian opposed to a non-Bayesian strategy on predictive performance and model confidence was evaluated with a CNN featuring the same architecture but using standard dropout (1Ch-NoMC). All CNNs were implemented in Keras with Tensorflow backend (Chollet & others, 2015). Code is made available at Github for reproducibility (https://github.com/liherz/stroke_classification).

For model fitting, we trained with respect to the binary cross-entropy loss (Goodfellow, et al., 2016)

$$L = -\frac{1}{N}\sum_{i=1}^{N} y_i \log(p_1(x_i)) + (1-y_i)\log\,(p_0(x_i)) \tag{1}$$

with $y_i$ the true label of image $x_i$ (equal to 1 for stroke and 0 for no-stroke), $p_1(x_i)$ and $p_0(x_i)$ the predicted probabilities for stroke and no-stroke and $N$ the total number of training data. The optimization algorithm applied was Adam (Kingma & Lei Ba, 2014) with default settings of the Keras implementation including a learning rate of 0.001. We trained for 400 epochs using a mini-batch size of 64. To avoid overfitting, early stopping was performed (Prechelt, 2012). That is, we retrospectively searched for the epoch in which the validation loss was minimal and considered the corresponding parameters as the trained CNN. For further regularization, the data set was extended artificially using real-time data augmentation (Wang & Perez, 2017). To catch the heterogeneous shapes, spatial locations and sizes of stroke lesions, the 2D images were randomly zoomed with a factor of 50-150%, shifted along the width and height within a range of 20%, sheared with a shearing angle of 20° degrees and rotated up to a degree of 20°. For that, we applied the ImageDataGenerator in Keras (Chollet & others, 2015).

## 2.3 MC Dropout and Uncertainty Measures

In a conventional, non-Bayesian CNN as obtained with standard dropout (Srivastava, et al., 2014), the parameters $\omega$ of the network are single numbers, which yield point predictions $\hat{p}_c$ in terms of probabilities that an input image $x$ belongs to a class $c, c \in \{1,..,C\}$. These probabilities are calculated within the final network layer when applying the softmax function to each of the $c$ output units $z_c$, where

$$\hat{p}_c = \text{softmax}\,(z_c) = \frac{\exp\,(z_c)}{\sum_{j=1}^{C} \exp\,(z_j)}. \quad (2)$$

During training, the network parameters are learned by optimizing the loss function on the training data. At test time, the network yields class predictions for a new test image $x^*$; in our application, those are two probabilities, one for stroke and one for no-stroke.

In contrast, in a Bayesian setting, we place a prior distribution $f(\omega)$ over the model parameters $\omega$ and learn the posterior distribution $f(\omega|(x,y)) = \frac{f(X|\omega)f(\omega)}{\int f(X|\omega)f(\omega)d\omega}$ based on the training data with respect to the loss function. Then, the goal is to obtain a posterior predictive distribution for an output $y^*$ of a new test sample $x^*$, with

$$f(y^*|x^*,(x,y)) = \int f(y^*|x^*,\omega)\, f(\omega|(x,y))d\omega \quad (3)$$

where $f(y^*|x^*,\omega)$ is the likelihood function and $f(\omega|(x,y))$ the posterior distribution (Held & Sabanés Bové, 2014). Opposed to the non-Bayesian approach, this posterior predictive defines a whole distribution over probabilities that an image represents a stroke respectively no-stroke, instead of yielding two predictions only (see Figure 3). In case of DL models, due to the high dimensional parameter space, there is no analytical solution to the posterior $f(\omega|(x,y))$. However, is has been shown that an approximating solution can be obtained with dropout (Gal & Ghahramani, 2016). Keeping dropout turned on at test time enables to sample parameter sets from the approximate posterior to acquire independent and identically distributed samples of the posterior predictive distribution (MC sampling). In practice, this is equivalent to passing the same test image $x^*$ through $t = 1, \dots, T$ stochastic dropout perturbations of the network (MC Dropout) yielding several possible outcome predictions (see Figure 3). The spread of the resulting predictive distribution indicates the uncertainty in the class prediction. For a detailed explanation and derivation of the method, we refer to (Gal & Ghahramani, 2016).

To summarize the predictive posterior, a point estimate for the predictive probability and corresponding uncertainty measures were determined. For prediction of a class probability, we averaged across the softmax probabilities $\hat{p}_{c,t}$ of the respective class $c = 1,..,C$ resulting from the multiple dropout runs $t, t = 1, \dots, T$. This yield mean probabilities

$$\bar{p}_c = \frac{1}{T}\sum_{t=1}^{T} \hat{p}_{c,t}\,. \quad (4)$$

Uncertainty was assessed using the established uncertainty measures (see Section 1.1), including the MC dropout variance, variation ratio, predictive entropy and the mutual information. In addition, we differentiated between epistemic and aleatoric uncertainty.

The MC dropout variance (Var) is the variance across the softmax probabilities $\hat{p}_{c,t}$ resulting from the multiple dropout runs $t, t = 1, \dots, T$ and is defined as the average across the variances for classes $c = 1, \dots, C$:

$$\text{Var} = \frac{1}{C}\sum_{c}\frac{1}{T}\sum_{t=1}^{T}(\hat{p}_{c,t} - \bar{p}_c)^2. \quad (5)$$

The variation ratio (VR) indicates how much the predictive distribution varies around the mode $m$, which is the class occurring most frequently within the $T$ network runs, and is calculated as

$$\text{VR} = 1 - \frac{n_m}{T}. \quad (6)$$

Here, $n_m = \sum_{t=1}^{T} \text{I}(\hat{y}_t = m)$ is the number of mode occurrences in the $T$ dropout runs with I the indicator function and $\hat{y}_t$ the predicted class in pass $t$ (equal to 1 if the stroke probability exceeds the threshold of 0.5; 0 otherwise).

The predictive entropy (PE) is based on the mean probabilities for classes $c = 1,..,C$ and captures the average amount of information comprised in the predictive distribution. It is defined as

$$\text{PE} = -\sum_{c} \bar{p}_c \log(\bar{p}_c). \tag{7}$$

The mutual information (MI) is obtained as a combination of the predictive entropy and an average across the MC dropout predictions multiplied by their logarithmic version:

$$\text{MI} = -\sum_{c} \bar{p}_c \log(\bar{p}_c) + \frac{1}{T}\sum_{c,t} \hat{p}_{c,t} \log{(\hat{p}_{c,t})}. \tag{8}$$

Epistemic and aleatoric uncertainty are estimated using the approach developed and evaluated in (Kwon, et al., 2020) and (Ryu, et al., 2019). Epistemic uncertainty captures the uncertainty in the model parameters and is equivalent to the MC dropout variance (s. Equation (5)):

$$\text{Epi} = \frac{1}{C}\sum_{c}\frac{1}{T}\sum_{t=1}^{T}(\hat{p}_{c,t} - \bar{p}_c)^2. \tag{9}$$

Aleatoric uncertainty models the uncertainty in the training data and is estimated as the average across the variances of the Bernoulli-distributed softmax probabilities $\hat{p}_{c,t}$ from the MC dropout run $t, t = 1, \ldots, T$ with

$$\text{Alea} = \frac{1}{T}\sum_{t=1}^{T}\hat{p}_{1,t}(1 - \hat{p}_{1,t}) \tag{10}$$

and $\hat{p}_{1,t}$ the softmax probability for stroke class $c = 1$.
In addition, instead of summarizing the predictive distribution with mean probabilities and uncertainty measures, we estimated the distribution for classes $c = 1, .., C$ using histogram counts. For that, the probability range $[0, 1]$ was binned in non-overlapping intervals of sizes 0.01 and the amount of predictions falling into the intervals was counted and normalized, yielding prediction bin values:

$$\hat{p}_{c,bin_j} = \frac{1}{T}\sum_{t=1}^{T} I(\hat{p}_{c,t} \in bin_j)\ , j = 1, \ldots, 100. \tag{11}$$

## 2.4 Aggregation models for patient-level predictions

Since diagnoses and treatment decisions are made on the patient-level, the image predictions of the CNNs were combined to a patient diagnosis. Therefore, we evaluated several aggregation methods expecting different image-level input values.
As baseline aggregation method, we defined the stroke class probability of a patient as the maximum over the image-level stroke class probabilities associated with the respective patient (Maximum method). If this prediction exceeded a threshold of 0.5, the patient was classified as stroke patient, otherwise as TIA. Although this method is straightforward, it has major drawbacks. First, the patient diagnosis relies on one single image. Second, the confidence of the image prediction is ignored and third, no uncertainty measure at the patient-level is obtained. To overcome the disadvantages, we developed aggregation models more sophisticated like Fully Connected Neural Networks (FC-NNs) and one-dimensional CNNs (1D-CNNs) with MC Dropout applied (see Figure 4). Those approaches expect measures from multiple images of a patient and provide uncertainty information at the patient-level. In addition, the aggregation methods incorporate the uncertainty information about the image predictions. This is motivated by the fact that high uncertainty correlates with erroneous predictions, which may lead to a learned down weighting of information coming from uncertain input measures.
The FC-NN comprises three hidden layers having eight units each, followed by a two-unit fully connected layer with softmax activation attached (see Figure 4 (a)). ReLU non-linearity and dropout layers (with dropout probability 0.3) were inserted right after each fully connected layer. Since FC-NNs expect a fixed number of inputs, we provided the five images, which most likely represented stroke lesions in terms of the five highest stroke class probabilities $P$ associated with a respective patient. Experiments, in which we processed more images, yielded no improvement in model performance but rather resulted in overfitting (data not shown). To integrate the uncertainty in the image predictions, the network was extended to a parallel structure (see Figure 4 (b)-(d)). Each pathway exhibits the same architecture as the FC-NN described above but expects for each of the five images the stroke class probabilities $P$ along with (i) the uncertainty measures $\text{Var}, \text{VR}, \text{MI}$ and $\text{PE}$ respectively (ii) the uncertainty measures for epistemic and aleatoric uncertainty (Epi, Alea). Furthermore, we provided (iii) an estimate of the predictive distribution of the stroke probabilities; the histogram counts. As we expected the influence of the inputs to be similar among the five input images, model parameters across pathways were shared. To learn the contribution of each image to the patient prediction, pathway outputs were concatenated and an eight-unit fully connected layer, a dropout and a two-unit fully connected layer with softmax activation were attached. Dropout probability was increased to 0.4 in (i) and (ii) and to 0.5 in (iii) in order to account for the higher number of trainable parameters.

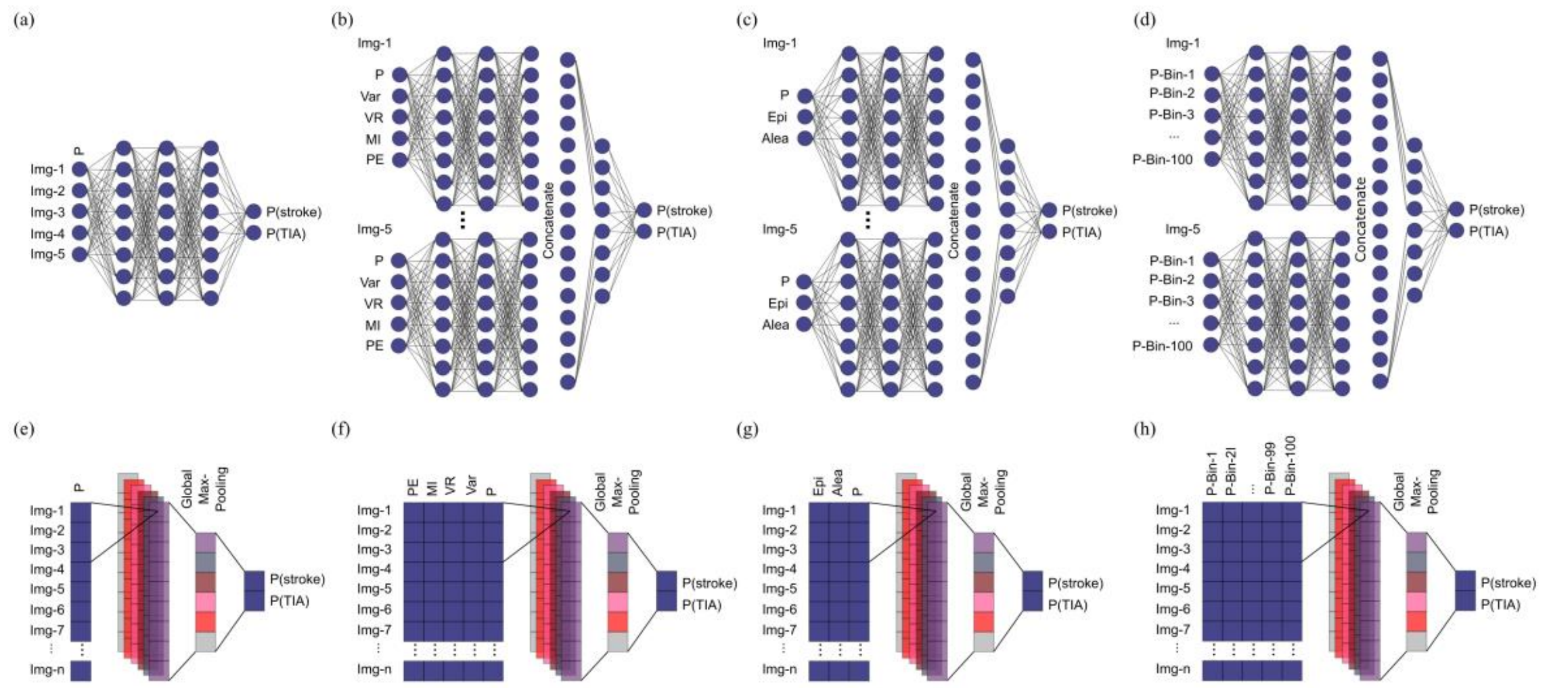

Figure 4. **Aggregation models to combine the image-level predictions to a patient-level diagnosis.** We developed fully connected neural networks (FC-NNs) (upper row) and 1D CNNs (lower row) and applied the networks to each set of image predictions resulting from one of the three CNNs. (a) A FC-NN expecting the five stroke class probabilities (P) of the five images, which most likely show a stroke lesion. (b) & (c) Parallel FC-NNs expecting the stroke class probabilities (P) of the five most likely stroke images and corresponding uncertainty measures (Var, VR, MI, PE respectively Epi, Alea). (d) A parallel FC-NN, expecting the histogram counts of the five images, which most likely show a stroke. (e) A 1D-CNN, taking in all stroke class probabilities (P) of all images n corresponding to a patient. (f) & (g) 1D-CNNs, additionally expecting the corresponding uncertainty measures. (h) A 1D-CNN, expecting the histogram counts of all images corresponding to a patient.

Unlike FC-NNs, 1D-CNNs can handle variable input sizes, which allowed us to consider all images of a patient for diagnosis. As input, the model expects the image-level stroke class probabilities of a patient. The 1D-CNN features one hidden layer with 16 units, a dropout and a global max-pooling layer followed by a two-unit fully connected layer with softmax activation attached (see Figure 4 (e)). ReLU non-linearity was used as activation function and dropout probabilities were fixed to 0.4. The filter-size was set to 3x1 pixels. Increasing the number of layers or the filter size showed no improvement (data not shown). As in the FC-NN, image-level prediction uncertainty was inserted in three ways. First, uncertainty measures as Var, VR, MI, PE and Epi, Alea were added to the input vector, which was extended to a second dimension (see Figure 4 (f) & (g)). Second, we replaced the probabilities and uncertainty measures with the histogram counts (see Figure 4 (h)). Here, the dropout rate was increased to 0.5. In addition, we trained simple and bi-directional Long short-term memory (LSTM) models. However, this increased the training time significantly while providing similar results (data not shown). FC-NNs and 1D-CNNs were trained for 200 epochs executing the Adam algorithm with Keras default settings and a batch size of 2. Binary cross-entropy loss was evaluated to measure classification performance. Similar to the procedure applied in the CNNs for image predictions, we retrospectively searched for minimal validation loss and used the parameters of the corresponding epoch for our final model. At test time, we applied dropout to obtain uncertainty information at the patient-level by running the same input through 500 dropout perturbations.

## 2.5 Training and Testing procedure

Five-fold cross validation was employed to make use of the whole data (see Figure 5). The data set was split into five test sets consisting of 102 and 103 patients, respectively. Within each of the five folds,

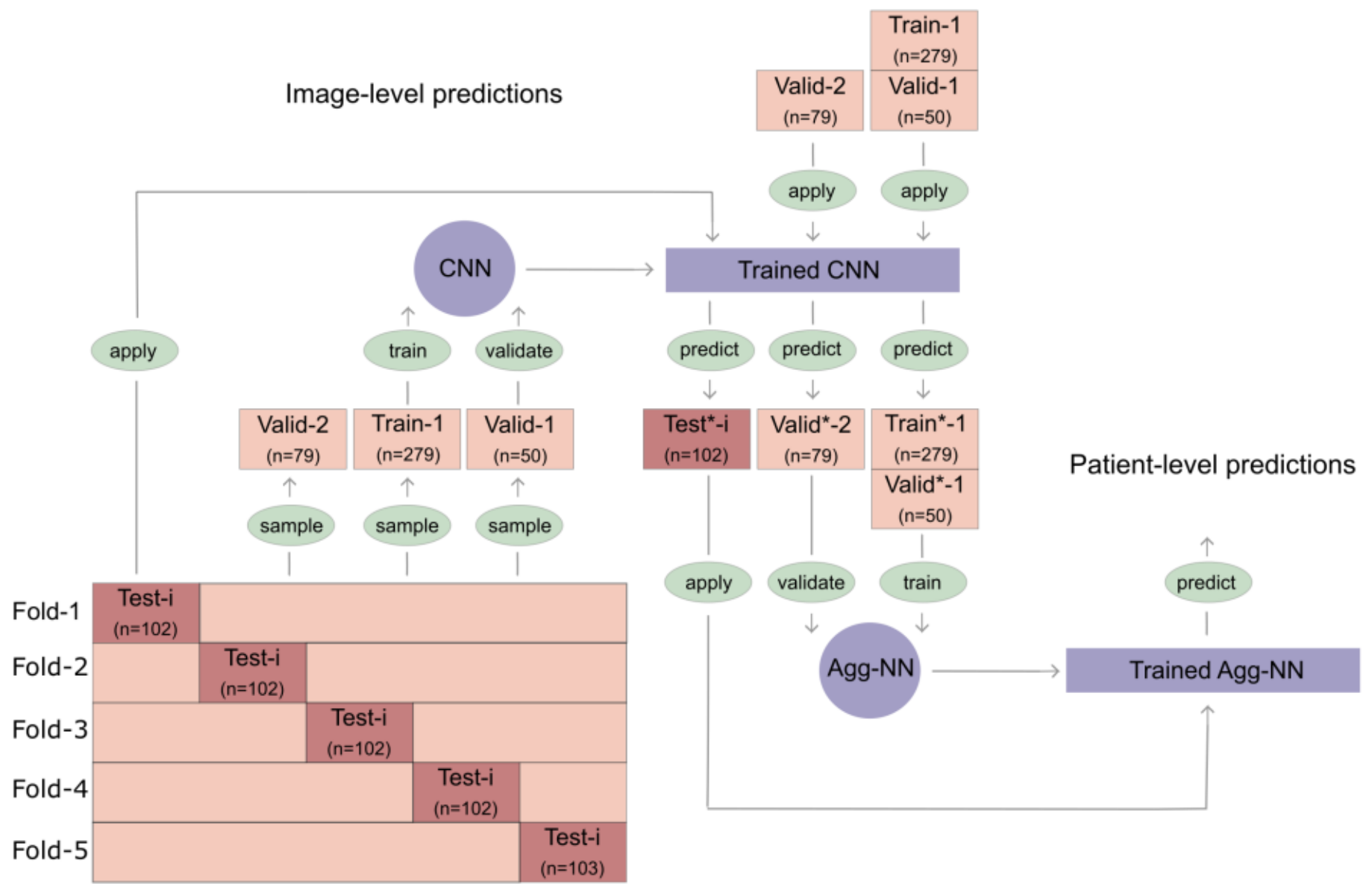

Figure 5. **Overview of the training and testing procedure with five-fold cross validation.** The data set was split in five subsets of equal sizes (Test-i). In each fold/run, patients who were not contained in Test-i, were randomly sampled to Train-1, Valid-1 and Valid-2. A CNN for image-level predictions was trained and validated on Train-1 and Valid-1, respectively. Then, the trained CNN was used to predict images in Train-1, Valid-1, Valid-2 and Test-i. Resulting image-level predictions in Train*-1, Valid*-1 and Valid*-2 were used to train and validate the aggregation models (see Figure 4). Image-level predictions in Test*-1 were obtained with the trained CNN and combined to a patient-level diagnosis using the trained aggregation models.

patients not contained in Test-i (i=1,..,5) were randomly sampled to Train-1, Valid-1 and Valid-2. For each set, we ensured that the percentage of stroke and TIA patients was similar. Subsequently, the sets were applied as follows for training, validation and testing:

CNNs for image predictions were trained and validated on Train-1 and Valid-1, respectively. Stroke images in Train-1 were duplicated to account for the imbalance among stroke and no-stroke images. Then, the trained CNNs were used to predict all images in Train-1, Valid-1 and Valid-2. Image predictions contained in Train*-1 and Valid*-1 were used to train the aggregation methods. Previously unseen image predictions in Valid*-2 were applied to validate the methods. Using the image predictions in Train*-1 and Valid*-1 to train the aggregation methods might cause some problems because those predictions could be higher in confidence than predictions in Valid*-2 and Test*-i as they were already seen by the CNN during training. However, we graphically compared the approximate predictive distributions of Train*-1, Valid*-1 and Valid*-2 and found no serious differences (data not shown). At test time, the previously unseen test data (Test-i) was passed through the trained CNNs. Then, the resulting image predictions in Test*-i were run through the trained aggregation methods to obtain patient-level predictions.

## 2.6 Evaluation metrics

Performance of the prediction models was evaluated in terms of discrimination and calibration (Held & Sabanés Bové, 2014). Discrimination measures how well the model predicts the outcome and is quantified with the accuracy along with 95% Wilson confidence intervals. Calibration measures the consensus between observed and predicted probabilities, taking into account the entire predictive distribution and not only the point predictions (Held & Sabanés Bové, 2014). For empirical assessment of calibration, we allocated stroke probabilities to I intervals ([0,0.05],…,[0.95,1]) with representative mean probabilities $\pi_1 = 0.025, \ldots, \pi_I = 0.975$ and calculated the number $n$ of events observed in interval $i$. The true probability $\bar{y}_i$ is defined as the percentage of true ischemic stroke events among all events in interval $i$. To visualize the calibration performance, calibration plots were considered. Furthermore, Sanders' calibration was used as a measure of calibration with $SC = \frac{\sum_i n_i(\bar{y}_i - \pi_i)^2}{N}$ and $N = \sum_{i=1}^{I} n_i$. In case of $SC = 0$, perfect calibration is achieved.

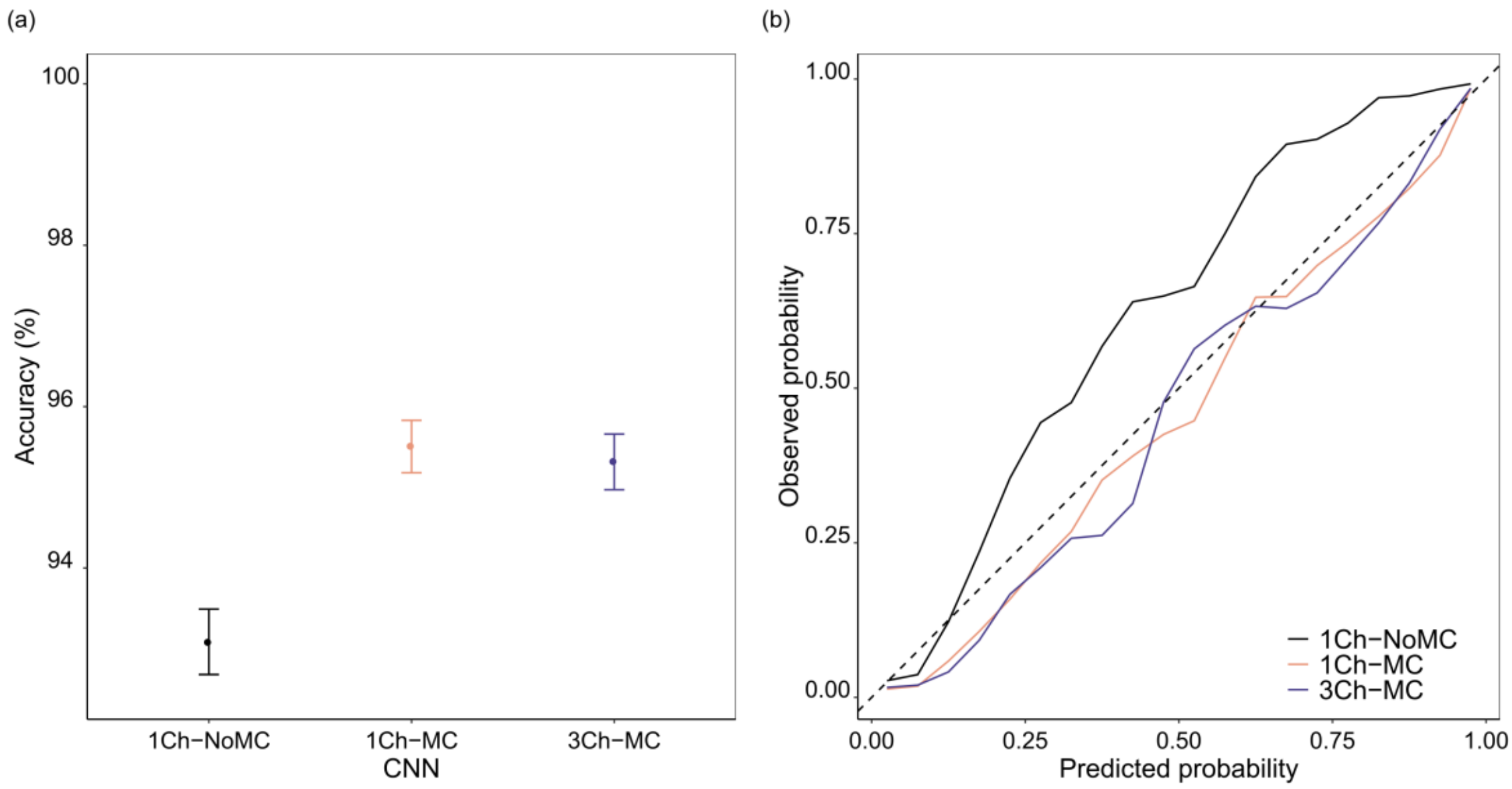

Figure 6. **Test set results of the three CNNs for the image-level predictions.** (a) Assessment of discrimination performance using the accuracy and 95% Wilson confidence intervals. (b) Calibration plot measuring the consensus between predicted and observed probabilities. In case of perfect calibration, curves would coincide with the main diagonal (dotted line).

# 3. Results

## 3.1 Test performance of image-level predictions

Discrimination performance of the three CNNs for image-level predictions is visualized in Figure 6 (a). In terms of accuracy, MC Dropout significantly improved performance. Unlike the 1Ch-MC and the 3Ch-MC, which achieved accuracies of 95.52% [95.18%, 95.83%] and 95.33% [94.97%, 95.66%], respectively, the 1Ch-NoMC with standard dropout reached a significantly lower accuracy of 93.09% [92.68%, 93.49%]. As all models feature the same architecture and were trained, validated and tested on the same set of patients, we conclude that the improvement is clearly due to MC Dropout. When we compared the 1Ch-MC (95.52% [95.18%, 95.83%]) to our proposed 3Ch-MC (95.33% [94.97%, 95.66%]), no difference in discrimination performance at the image-level was observed. However, as will be shown in Section 3.3, patient-level prediction performance was improved when considering neighboring information.

In terms of calibration, all three CNNs yielded well-calibrated predictions (see Figure 6 (b)). Nonetheless, the mean probabilities acquired with the MC Dropout networks achieved slightly lower Sanders' calibration scores of 0.004 (1Ch-MC) and 0.005 (3Ch-MC) opposed to the softmax predictions of the 1Ch-NoMC obtained with standard dropout (0.008).

## 3.2 Test performance of the derived uncertainty measures

As we aimed to integrate image-level prediction uncertainty into aggregation models for improved patient-level classification, the uncertainty measures summarized in Section 2.3 ($\mathrm{Var/Epi, VR, PE, MI, Alea}$) were expected to be able to distinguish erroneous from correct predictions. In the left panel of Figure 7, we considered the change in image-level prediction accuracy when removing images depending on the respective uncertainty measure. The right panel of Figure 7 visualizes Receiver Operating Characteristics (ROC) curves and reports the Area under the curve (AUC) statistic, to assess the power of the respective uncertainty measure to differentiate between correct and erroneous predictions. For the sake of simplicity, results of the 3Ch-MC are presented; the 1Ch-MC showed similar outcomes (see Appendix, Figure 1). Obviously, uncertainty-informed prediction removal increased accuracy clearly independent of the uncertainty measure used. As visualized in Figure 7 (a), accuracy improved by about 2% when removing only 5% of the images. In terms of AUC, the variance respectively the epistemic uncertainty measure (0.83 [0.81, 0.84]), the predictive entropy (0.83 [0.81, 0.85]), the mutual information (0.82 [0.80, 0.83]) and the aleatoric uncertainty measure (0.82 [0.81, 0.84]) performed similarly well. Only the variation ratio yielded results slightly worse (0.79 [0.78, 0.81]). Accordingly, our results confirmed previous findings that erroneous predictions correlate with high uncertainties (Dürr, et al., 2018; Leibig, et al., 2017; Nair, et al., 2020). Integrating uncertainty information into aggregation models (Figure 4 (b)-(d) and (f)-(h)) is therefore supposed to improve patient classification and appears reasonable. As all types of uncertainty measures performed similarly well in differentiating correct from erroneous predictions but do contain slightly different information, we decided to consider aggregation models including all metrics.

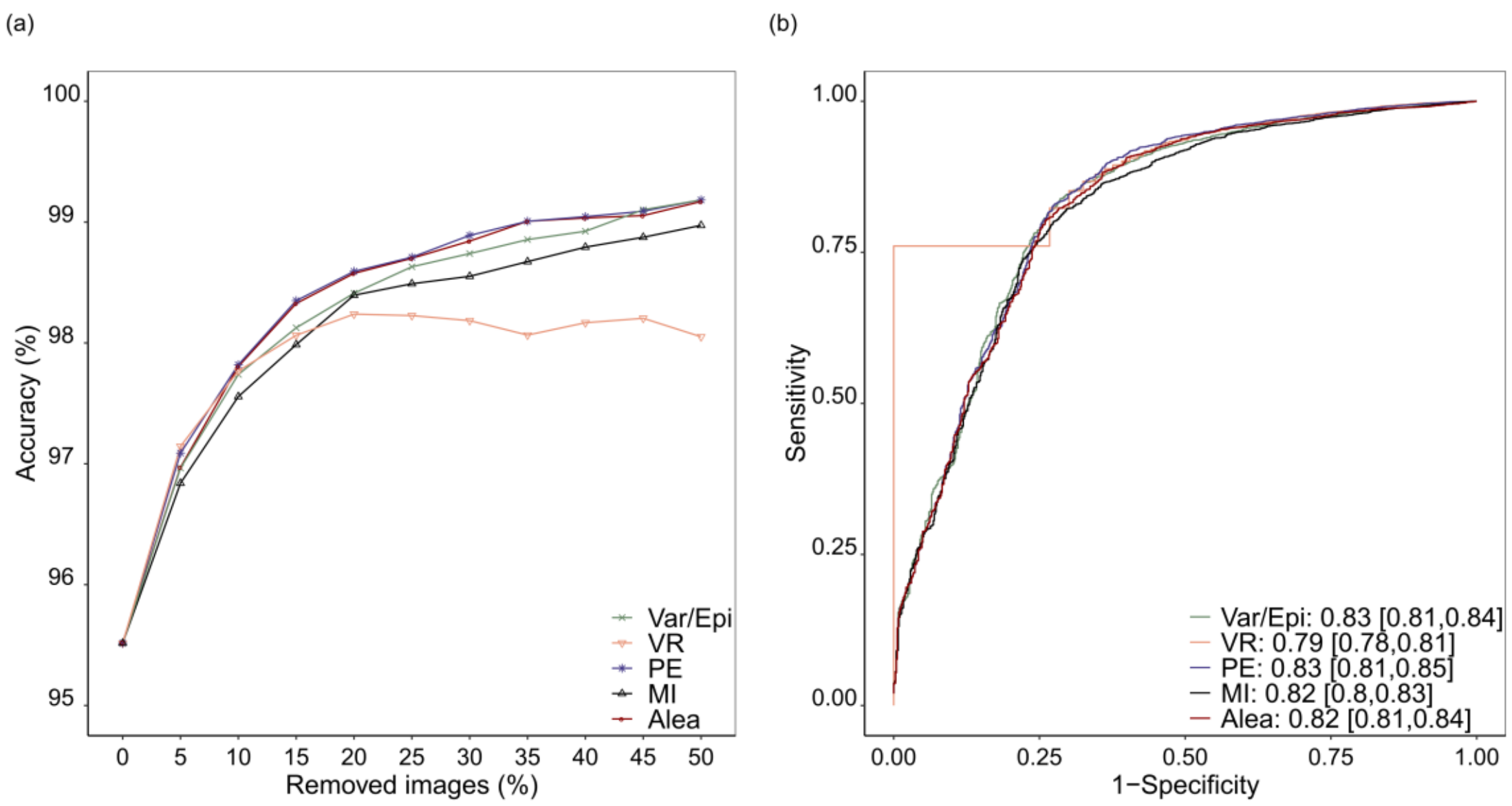

Figure 7. **The power of uncertainty measures to differentiate between correct and erroneous image-level predictions (3Ch-MC) evaluated on the test set.** (a) Uncertainty informed prediction removal according to the respective uncertainty measure (Variance/Epistemic ($Var/Epi$), Variation Ratio ($VR$), Predictive Entropy ($PE$), Mutual Information ($MI$), Aleatoric ($Alea$)). An increase in accuracy indicates a correlation between erroneous predictions and high uncertainty measures. (b) Receiver Operating Characteristics (ROC) curves and Areas under the Curve (AUCs). A value of one represents perfect discrimination.

### 3.3 Test performance of patient-level predictions

Discrimination performances of the aggregation models (Max, FC-NN, 1D-CNN) based upon the image predictions and uncertainty information acquired from the three CNNs (1Ch-NoMC, 1Ch-MC, 3Ch-MC) are summarized in Table 2. As already indicated in Section 3.1, using MC Dropout in deep CNNs improved classification performance clearly, also at the patient-level. Aggregation approaches based on softmax predictions of the 1Ch-NoMC, achieved in general 1-2% lower accuracies compared to methods utilizing image predictions of the 1Ch-MC (Max: 90.22% [87.33%, 92.50%] vs. 91.59% [88.86%, 93.69%]; FC-NN: 92.75% [90.16%, 94.69%] vs. 93.53% [91.05%, 95.36%]; 1D-CNN: 92.37% [89.74%, 94.37%] vs. 92.37% [89.74%, 94.37%]). Those results emphasize that MC dropout does not only provide uncertainty information but also improves classification performance. The accuracy of the aggregation methods executing image predictions of the 3Ch-MC increased further by 1-2% opposed to the 1Ch-MC. This indicates an advantageous effect on predictive performance when considering neighboring images for feature learning in the setting of a particular 3D data structure.

Regarding the aggregation models, the maximum method performed worst. Combining predictions of multiple or all images of a patient within a FC-NN or 1D-CNN, improved classification performance clearly. This is especially seen when using the predictions of the CNNs applying MC Dropout, the 1Ch-MC (91.59% [88.86%, 93.69%] vs. 93.53% [91.05%, 95.36%] in the FC-NN and 92.37% [89.74%, 94.37%] in the 1D-CNN) and the 3Ch-MC (93.15% [90.62%, 95.03%] vs. 95.69 [93.56, 97.13] in the FC-NN and 95.30% [93.11%, 96.82%] in the 1D-CNN). Strikingly, additionally providing uncertainty information in terms of uncertainty measures or an estimate of the predictive distribution yielded no valuable improvement in discrimination performance. Yet, the best results were obtained with the FC-NN using the uncertainty estimates for epistemic and aleatoric uncertainty as additional inputs (95.89% [93.79%, 97.29%]). However, there were no significant differences neither between FC-NNs and 1D-CNNs nor between predictions and predictions with uncertainty information as input.
In terms of calibration, the maximum method yielded the worst results, but all Sander's calibration scores approached zero indicating well-calibrated predictions (see Appendix, Table 1).
Applying MC Dropout in our aggregation models (FC-NN and 1D-CNN), additionally enabled to evaluate the models' uncertainty about patient-level predictions (see Figure 8). Again, correct predictions were supposed to be associated with high confidence measures (values close to zero); for erroneous predictions, the opposite held. For the sake of simplicity, uncertainty is represented using the predictive entropy, which appeared to be one of the best measures to differentiate between correct and erroneous predictions (see Figure 7). Yet, the remaining measures yielded similar results (see Appendix, Figures 2-5). In addition, we consider the 3Ch-MC only; results for the 1Ch-MC were similar (data not shown). As visualized in the upper panel of Figure 8, uncertainty informed prediction removal enables to improve the patient-level performance for all aggregation models. However, the models integrating uncertainty information about image predictions appear to perform better in that task. This is confirmed in particular when considering the AUC values of the ROC curves in the lower panels of Figure 8. The AUCs of the models integrating image-level uncertainties are increased compared to the models using the predictions only, indicating a better discrimination between erroneous and correct predictions. Here, the aggregation model incorporating the uncertainty measures variation ratio, predictive entropy, mutual information and variance yielded the best results.

Table 2. **Test set results at the patient-level.** Discrimination performance of the aggregation approaches (columns) with different input features (Predictions (P), Predictions and uncertainty measures (P + VR, PE, MI, Var, Epi, Alea), Histogram counts (Hist counts)) from the three CNNs (rows). The best results are highlighted in bold.

| | **Max** | **FC-NN** | | | | **1D-CNN** | | | |
|---|---|---|---|---|---|---|---|---|---|
| | | Predictions (P) | P + VR, PE, MI, Var | P + Epi, Alea | Hist counts | Predictions (P) | P + VR, PE, MI, Var | P + Epi, Alea | Hist counts |
| 1Ch-NoMC | 90.22 [87.33, 92.50] | **92.75 [90.16, 94.69]** | | | | 92.37 [89.74, 94.37] | | | |
| 1Ch-MC | 91.59 [88.86, 93.69] | 93.53 [91.05, 95.36] | 93.53 [91.05, 95.36] | 94.31 [91.95, 96.01] | 93.14 [90.61, 95.02] | 92.37 [89.74, 94.37] | **94.32 [91.97, 96.02]** | **94.32 [91.97, 96.02]** | 93.93 [91.52, 95.69] |
| 3Ch-MC | 93.15 [90.62, 95.03] | 95.69 [93.56, 97.13] | 94.90 [92.64, 96.50] | **95.89 [93.79, 97.29]** | 94.31 [91.95, 96.01] | 95.30 [93.11, 96.82] | 95.30 [93.11, 96.82] | 95.11 [92.88, 96.66] | 93.74 [91.29, 95.53] |

## 4. Discussion & Conclusion

In this paper, we demonstrate that integrating uncertainty information in deep neural network models is essential to achieve a high performance for stroke classification and to assess the reliability of individual predictions. By comparing a Bayesian CNN to its non-Bayesian counterpart, we showed advantageous effects of MC dropout on prediction performance. As MC dropout can easily be applied to existing network architectures (Gal & Ghahramani, 2016), it should be considered in all medical image application tasks not only for uncertainty estimation but also for performance improvement.

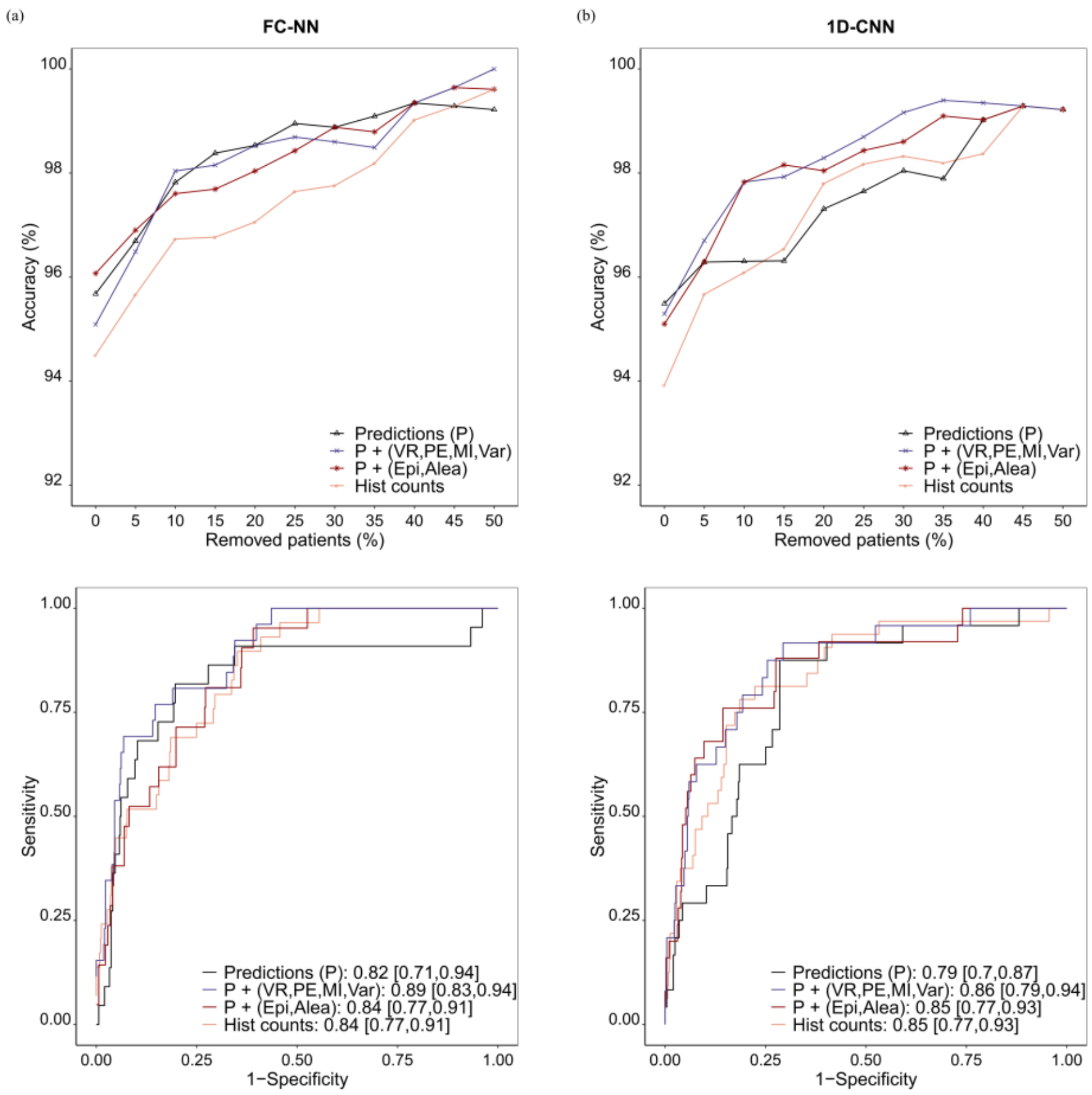

Figure 8. **Uncertainty in terms of the predictive entropy in patient-level predictions.** The two columns represent the results for the (a) FC-NNs and the (b) 1D-CNNs as aggregation models based on image-level predictions of the 3Ch-MC. The upper row shows the performance improvement when removing uncertain patient-level predictions. The ability of the different aggregation methods to differentiate between correct and erroneous patient-level predictions based on the uncertainty measure is summarized in the lower rows using ROC curves and corresponding AUC values.

The importance of integrating 3D information in the analysis of ischemic stroke patients and MRI data in general was highlighted when we compared CNN models with and without neighboring information to each other (3Ch-MC vs. 1Ch-MC). Although the image-level predictions of the proposed 3Ch-MC did not benefit from integrating neighboring images, that information was highly influential when combining the image predictions to patient diagnoses. Aggregation methods based on image predictions of the 3Ch-MC clearly outperformed the models incorporating image predictions of the 1Ch-MC at the patient-level.

We further confirmed previous findings that uncertainty estimates derived from the predictive distribution enable to differentiate erroneous from correct predictions and can therefore be used to filter false classifications. We considered all established uncertainty measures and in addition distinguished between epistemic and aleatoric uncertainty. Besides the variation ratio, all uncertainty measures considered (variance/epistemic uncertainty measure, predictive entropy, mutual information, aleatoric uncertainty measure) appeared to perform similarly well in detecting false predictions. Yet, the predictive entropy and the measure for aleatoric uncertainty slightly outperformed the other estimates. In practice, we can make use of that correlation between erroneous predictions and high uncertainties, by for example, holding uncertain cases back for additional inspection instead of predicting them. That may enable to find unknown or ambiguous cases to which a physician could consult further tools and experts.

To the best of our knowledge, we proposed and assessed one of the first neural network-based aggregation methods to combine a variable number of image predictions to patient diagnoses while simultaneously considering the image uncertainties and providing a measure of uncertainty at the patient level. In practice, such aggregation methods are frequently necessary, for instance, when combining voxel to lesion predictions, patch-level to image-level predictions and image-level to patient-level predictions as in our application setting. The benefit of integrating the image uncertainties was not observable in a performance improvement, but in an increased ability of the models to differentiate between correct and erroneous predictions. That is, integrating uncertainty information about image predictions enables to improve filtering of false patient diagnoses, which is highly important in practice. While the primary interest of our research was not on the development of neural network architectures but rather on integrating uncertainty in the common analysis procedure, our models achieved state-of-the art results with accuracies of more than 95% at the image- and the patient-level. In future applications, the CNN and the aggregation models may be combined to diagnose patients in an end-to-end fashion as for example done in (Patel, et al., 2019), to improve performance. Furthermore, incorporating more than only the neighboring images for feature learning and patient data when combining the image-predictions to patient diagnoses may be of additional interest.
In conclusion, this study highlights the importance of integrating uncertainty information not only in deep neural network architectures but also in aggregation models. Our proposed methods can directly be applied to other medical image classification tasks and may help to easily improve performance and reliability. This is particularly valuable in medical diagnosis and treatment, to alert the responsible physician using automated classification methods and facing potentially novel or unusual situations.

## Funding

This work was supported by the Swiss National Science Foundation (SNSF).

## Address correspondence to

Beate Sick (beate.sick@uzh.ch, sick@zhaw.ch) and Lisa Herzog (lisa.herzog@uzh.ch, hezo@zhaw.ch)

## References

Ayhan, M. S. & Berens, P., 2018. Test-time data augmentation for estimation of heteroscedastic aleatoric uncertainty in deep neural networks. *Medical Imaging with Deep Learning,* pp. 1-9.

Baird, A. E. & Warach, S., 1998. Magnetic Resonance Imaging of Acute Stroke. *Journal of Cerebral Blood Flow and Metabolism,* Volume 18, pp. 583-609.

Bernal, J. et al., 2019. Deep convolutional neural networks for brain image analysis on magnetic resonance imaging: a review. *Artificial Intelligence in Medicine,* Volume 95, pp. 64-81.

Chilla, G. S., Tan, C. H., Xu, C. & Poh, C. L., 2015. Diffusion weighted magnetic resonance imaging and its recent trend-a survey. *Quantitative imaging in medicine and surgery,* Volume 5, pp. 407-422.

Chollet, F. & others, 2015. *Keras.* [Online] Available at: https://keras.io

Dürr, O. et al., 2018. Know When You Don't Know: A Robust Deep Learning Approach in the Presence of Unknown Phenotypes. *Assay and Drug Development Technologies,* Volume 16, pp. 343-349.

Feng, R., Badgeley, M., Mocco, J. & Oermann, E. K., 2018. Deep learning guided stroke management: a review of clinical applications. *Journal of NeuroInterventional Surgery,* Volume 10, pp. 358-362.

Gal, Y., 2017. *Uncertainty in Deep Learning,* University of Cambridge: Department of Engineering.

Gal, Y. & Ghahramani, Z., 2016. Dropout as a Bayesian approximation: representing model uncertainty in deep learning. *International Conference on Machine Learning,* Volume 48.

Goodfellow, I., Bengio, Y. & Courville, A., 2016. *Deep Learning.* Cambridge: MIT Press.

Havaei, M. et al., 2017. Brain tumor segmentation with Deep Neural Networks. *Medical Image Analysis,* Volume 35, pp. 18-31.

Held, L. & Sabanés Bové, D., 2014. *Applied Statistical Inference.* 1 ed. Berlin Heidelberg: Springer-Verlag.

Ioffe, S. & Szegedy, C., 2015. Batch normalization: Accelerating deep network training by reducing internal covariate shift. *arXiv:1502.03167.*

Islam, J. & Zhang, Y., 2017. An ensemble of deep convolutional neural networks for Alzheimer's disease detection and classification. *NIPS 2017 Workshop on Machine Learning for Health.*

Kamnitsas, K. et al., 2015. Multi-Scale 3D Convolutional Neural Networksfor Lesion Segmentation in Brain MRI. *MICCAI Ischemic Stroke Lesion Segmentation Challenge.*

Kamnitsas, K. et al., 2016. Efficient multi-scale 3D CNN with fully connected CRF for accurate brain lesion segmentation. *Medical Image Analysis,* Volume 36, pp. 61-78.

Kendall, A., Badrinarayanan, V. & Cipolla, R., 2017. Bayesian SegNet: Model Uncertainty in Deep Convolutional Encoder-Decoder Architectures for Scene Understanding. *British Machine Vision Conference.*

Kendall, A. & Gal, Y., 2017. What uncertainties do we need in Bayesian deep learning?. *31st Conference on Neural Information Processing System (NIPS).*

Kingma, D. P. & Lei Ba, J., 2014. Adam: A Method for Stochastic Optimization. *International Conference on Learning Representations.*

Kwon, Y., Won, J.-H., Kim, B. J. & Paik, M. C., 2020. Uncertainty quantification using Bayesian neural networks in classification: Application to biomedical image segmentation. *Computational Statistics & Data Analysis,* Volume 142.

Lai, M., 2015. Deep learning for medical image segementation. *arXiv:1505.020.*

Leibig, C. et al., 2017. Leveraging uncertainty information from deep neural networks for disease detection. *Scientific Reports,* Volume 7.

Lisowska, A., Beveridge, E., Muir, K. & Poole, I., 2017. Thrombus Detection in CT Brain Scans using a Convolutional Neural Network. *Proceedings of the 10th International Joint Conference on Biomedical Engineering Systems and Technologie,* pp. 24-33.

Nair, T., Precup, D., Arnold, D. L. & Arbel, T., 2020. Exploring Uncertainty Measures in Deep Networks for Multiple Sclerosis Lesion Detection and Segmentation. *Medical Image Analysis,* Volume 59.

Nair, V. & Hinton, G. E., 2010. Rectified linear units improve restricted Boltzmann machines. *Proceedings of the 27th International Conference on International Conference on Machine Learning,* pp. 807-814.

Ozdemir, O., Woodward, B. & Berlin, A. A., 2017. Propagating Uncertainty in Multi-Stage Bayesian Convolutional Neural Networks with Application to Pulmonary Nodule Detection. *arXiv:1712.00497v1.*

Patel, A. et al., 2019. Image Level Training and Prediction: Intracranial Hemorrhage Identification in 3D Non-Contrast CT. *IEEE Access,* Volume 7, pp. 1-1.

Payan, A. & Montana, G., 2015. Predicting Alzheimer's disease: a neuroimaging study with 3D convolutional neural networks. *ICPRAM 2015 - 4th International Conference on Pattern Recognition Applications and Methods, Proceedings.*

Pinto, A. et al., 2018. Stroke Lesion Outcome Prediction Based on MRI Imaging Combined With Clinical Information. *Frontiers in Neurology,* Volume 9, p. 1060.

Prechelt, L., 2012. Early Stopping - but when?. In: *Montavon G., Orr G.B., Müller KR. (eds) Neural Networks: Tricks of the Trade. Lecture Notes in Computer Science.* Berlin, Heidelberg: Springer.

Purves, D. et al., 2004. *Neuroscience.* 3 ed. U.S.A.: Sinauer Associates.

Rajpurkar, P. et al., 2017. CheXNet: Radiologist-Level Pneumonia Detection on Chest X-Rays with Deep Learning. *arXiv:1711.05225.*

Roy, A. G., Conjeti, S., Navab, N. & Wachinger, C., 2018. Inherent brain segmentation quality control from fully convnet Monte Carlo sampling. *Interenational Conference on Medical Image Computing and Computer-Assisted Intervention,* pp. 664-672.

Ryu, S., Kwon, Y. & Kim, W. Y., 2019. A Bayesian graph convolutional network for reliable prediction of molecular properties with uncertainty quantification. *Chemical Science,* Volume 36, pp. 8438-8446.

Saver, J. L., 2015. Time is Brain - Quantified. *Stroke,* Volume 37, pp. 263-266.

Setio, A. et al., 2016. Pulmonary Nodule Detection in CT Images: False Positive Reduction Using Multi-View Convolutional Networks. *IEEE Transactions on Medical Imaging,* Volume 35.

Simonyan, K. & Zisserman, A., 2014. Very Deep Convolutional Networks for Large-Scale Image Recognition. *arXiv 1409.1556.*

Srivastava, N. et al., 2014. Dropout: A Simple Way to Prevent Neural Networks from Overfitting. *Journal of Machine Learning Research,* Volume 15, pp. 1929-1958.

Tousignant, A. et al., 2019. Prediction of Disease Progression in Multiple Sclerosis Patients using Deep Learning Analysis of MRI Data. *Proceedings of The 2nd International Conference on Medical Imaging with Deep Learning,* Volume 102, pp. 483-492.

Wang, G. et al., 2019. Aleatoric uncertainty estimation with test-time augmentation for medical image segmentation with convolutional neural networks. *NeuroComputing,* Volume 338, pp. 34-45.

Wang, J. & Perez, L., 2017. The Effectiveness of Data Augmentation in Image Classification using Deep Learning. *arXiv:1712.04621.*

Zhao, G. et al., 2018. Bayesian convolutional neural network based MRI brain extraction on nonhuman primates. *NeuroImage,* Volume 175, pp. 32-44.

# Appendix

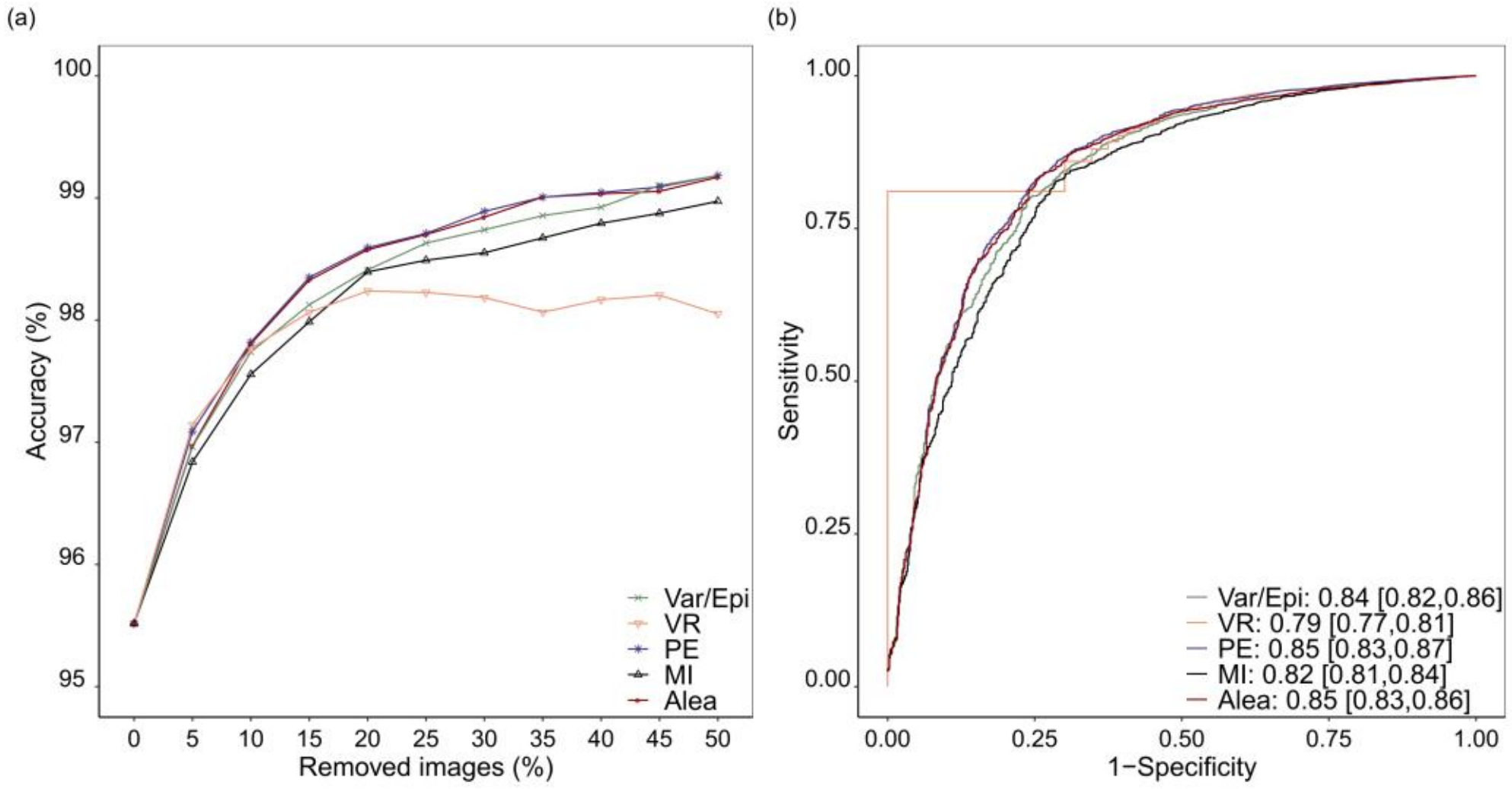

Figure 1: **The power of uncertainty measures to differentiate between correct and erroneous image-level predictions (1Ch-MC) evaluated on the test set.** (a) Uncertainty informed prediction removal according to the respective uncertainty measure (Variance/Epistemic ($Var/Epi$), Variation Ratio ($VR$), Predictive Entropy ($PE$), Mutual Information ($MI$), Aleatoric ($Alea$)). An increase in accuracy indicates a correlation between erroneous predictions and high uncertainty measures. (b) Receiver Operating Characteristics (ROC) curves and Areas under the Curve (AUCs). A value of one represents perfect discrimination.

Table 1. **Sanders' calibration scores.** Sander's calibration scores for different combinations of image-level predictions (rows) and aggregation methods (columns).

| | **Max** | **FC-NN** | | | | **1D-CNN** | | | |
|---|---|---|---|---|---|---|---|---|---|
| | | Predictions (P) | P + VR, PE, MI, Var | P + Epi, Alea | Hist counts | Predictions (P) | P + VR, PE, MI, Var | P + Epi, Alea | Hist counts |
| 1Ch-NoMC | 0.009 | 0.009 | | | | 0.01 | | | |
| 1Ch-MC | 0.05 | 0.009 | 0.009 | 0.007 | 0.01 | 0.01 | 0.01 | 0.006 | 0.008 |
| 3Ch-MC | 0.03 | 0.01 | 0.009 | 0.007 | 0.01 | 0.007 | 0.007 | 0.007 | 0.01 |

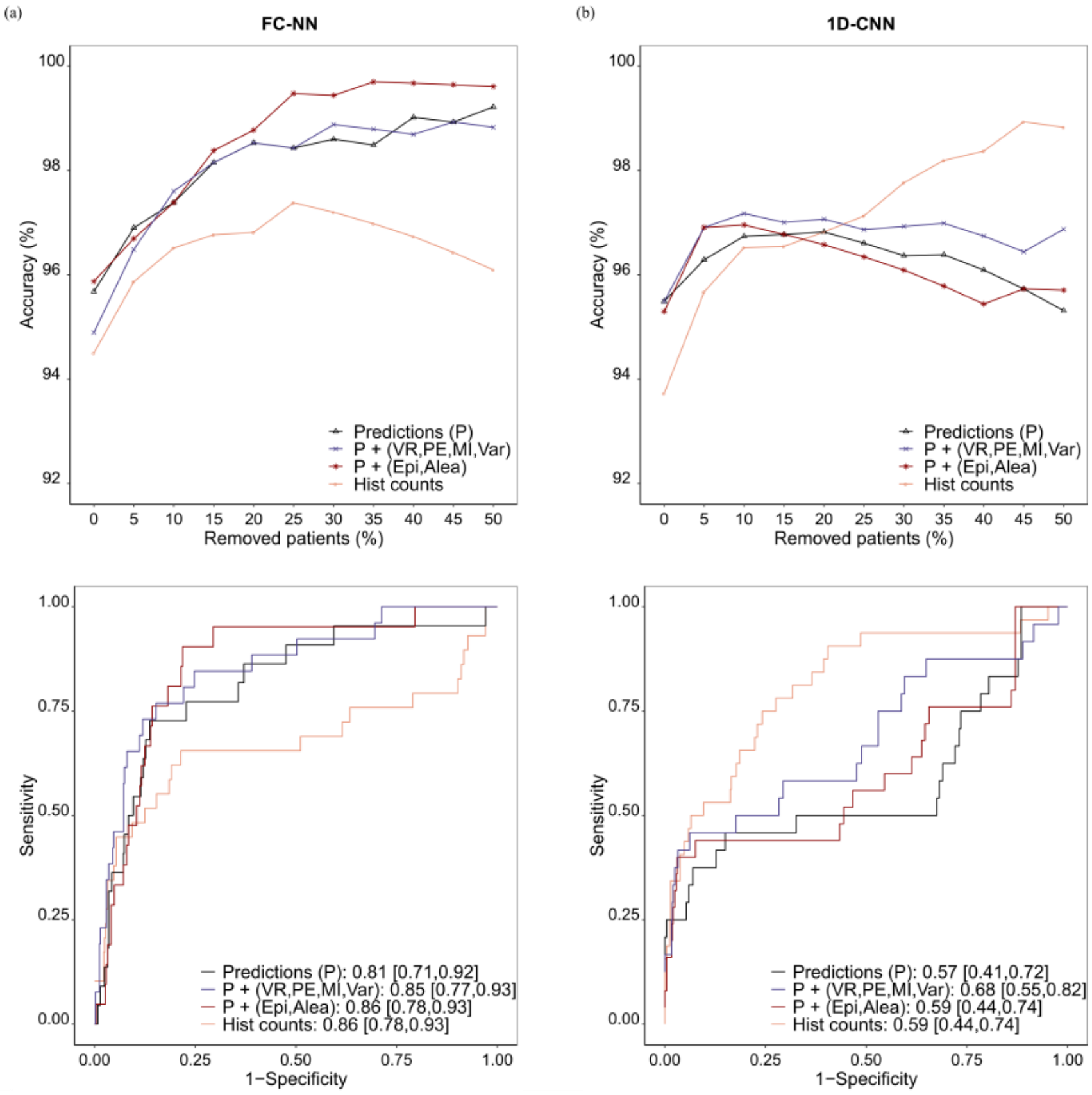

Figure 2. **Uncertainty in terms of the variance/epistemic uncertainty in patient-level predictions.** The two columns represent the results for the (a) FC-NNs and the (b) 1D-CNNs as aggregation models based on image-level predictions of the 3Ch-MC. The upper row shows the performance improvement when removing uncertain patient-level predictions. The ability of the different aggregation methods to differentiate between correct and erroneous patient-level predictions based on the uncertainty measure is summarized in the lower rows using ROC curves and corresponding AUC values.

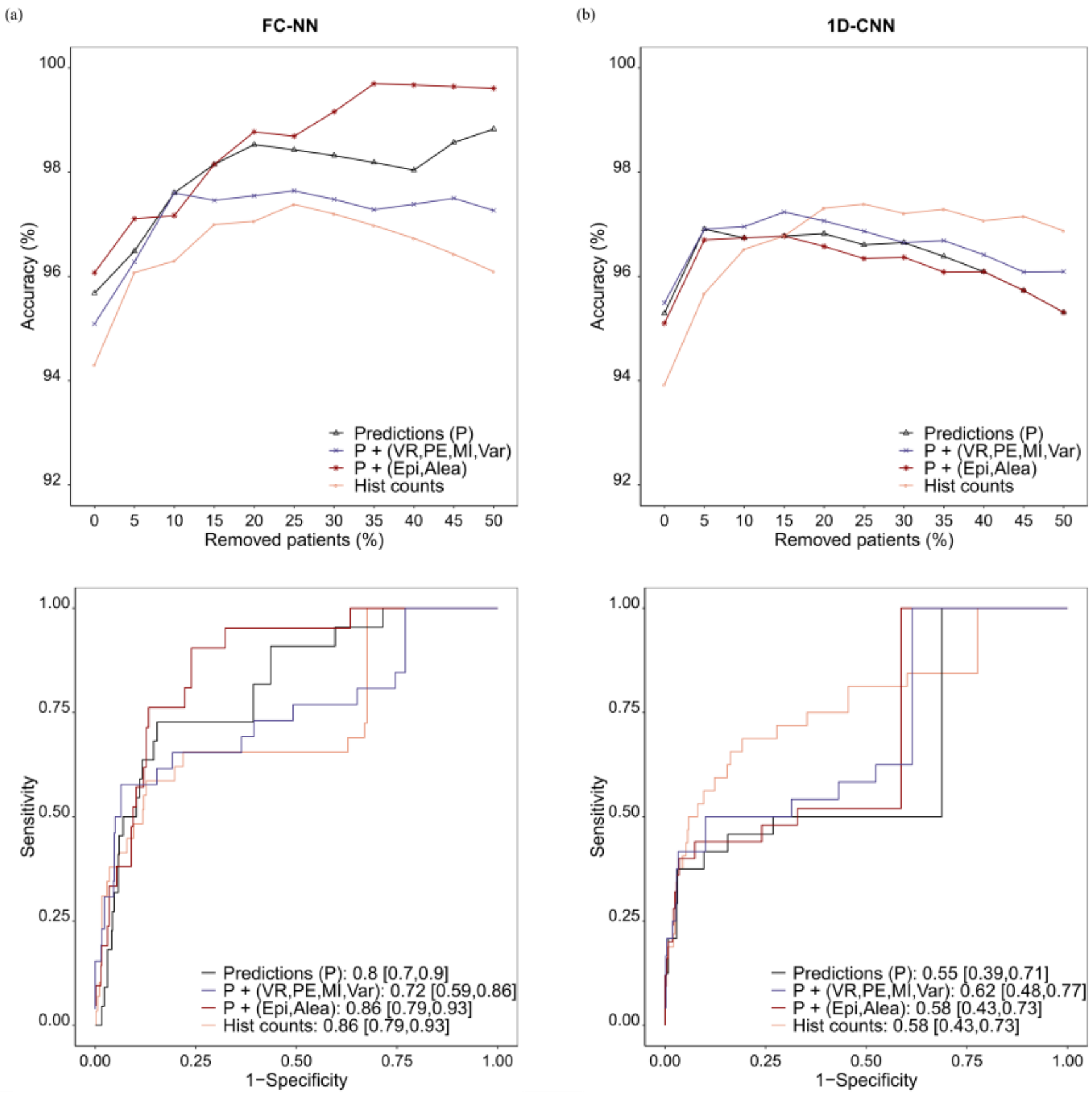

Figure 3. **Uncertainty in terms of the variation ratio in patient-level predictions.** The two columns represent the results for the (a) FC-NNs and the (b) 1D-CNNs as aggregation models based on image-level predictions of the 3Ch-MC. The upper row shows the performance improvement when removing uncertain patient-level predictions. The ability of the different aggregation methods to differentiate between correct and erroneous patient-level predictions based on the uncertainty measure is summarized in the lower rows using ROC curves and corresponding AUC values.

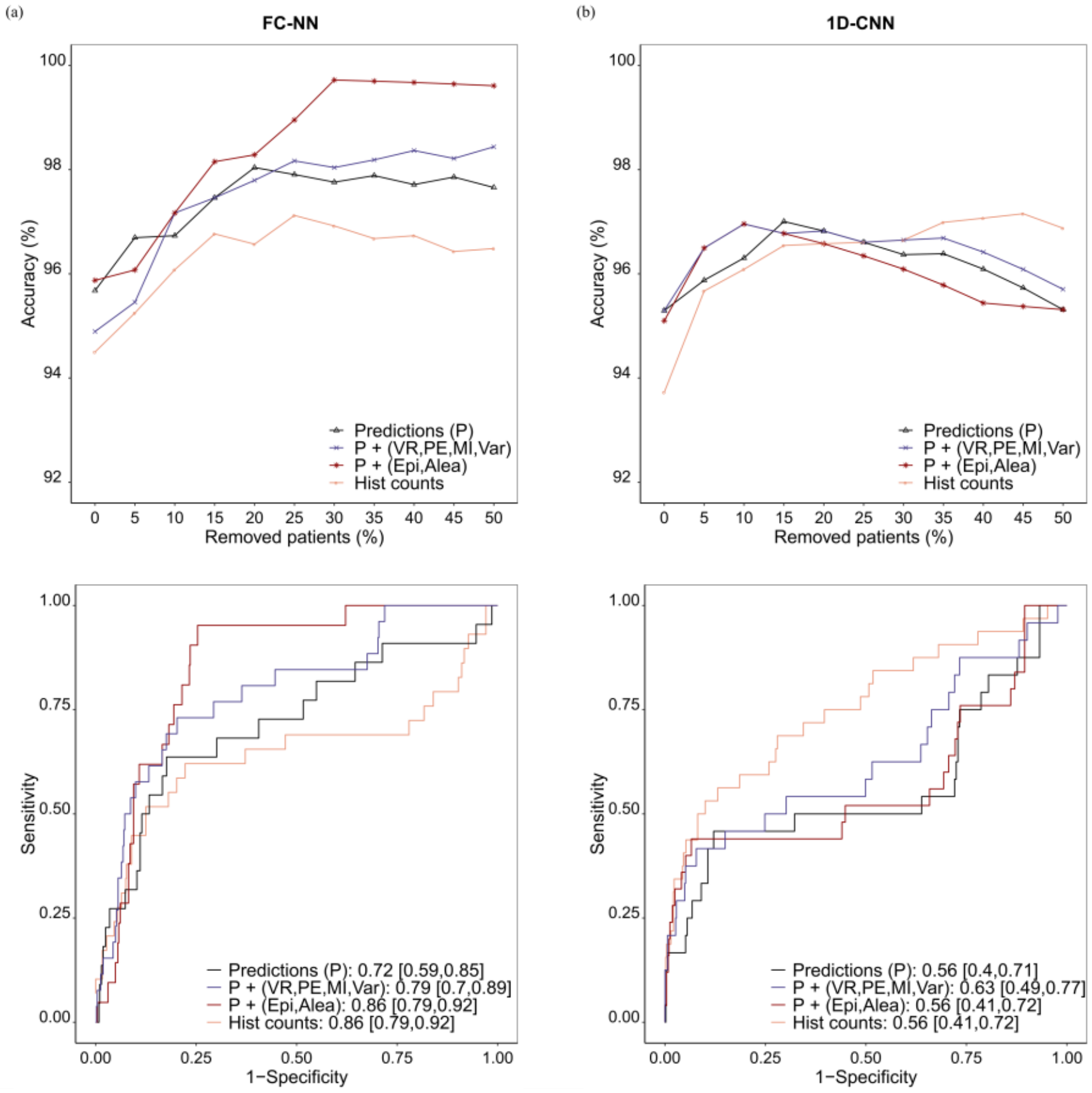

Figure 4. **Uncertainty in terms of the mutual information in patient-level predictions.** The two columns represent the results for the (a) FC-NNs and the (b) 1D-CNNs as aggregation models based on image-level predictions of the 3Ch-MC. The upper row shows the performance improvement when removing uncertain patient-level predictions. The ability of the different aggregation methods to differentiate between correct and erroneous patient-level predictions based on the uncertainty measure is summarized in the lower rows using ROC curves and corresponding AUC values.

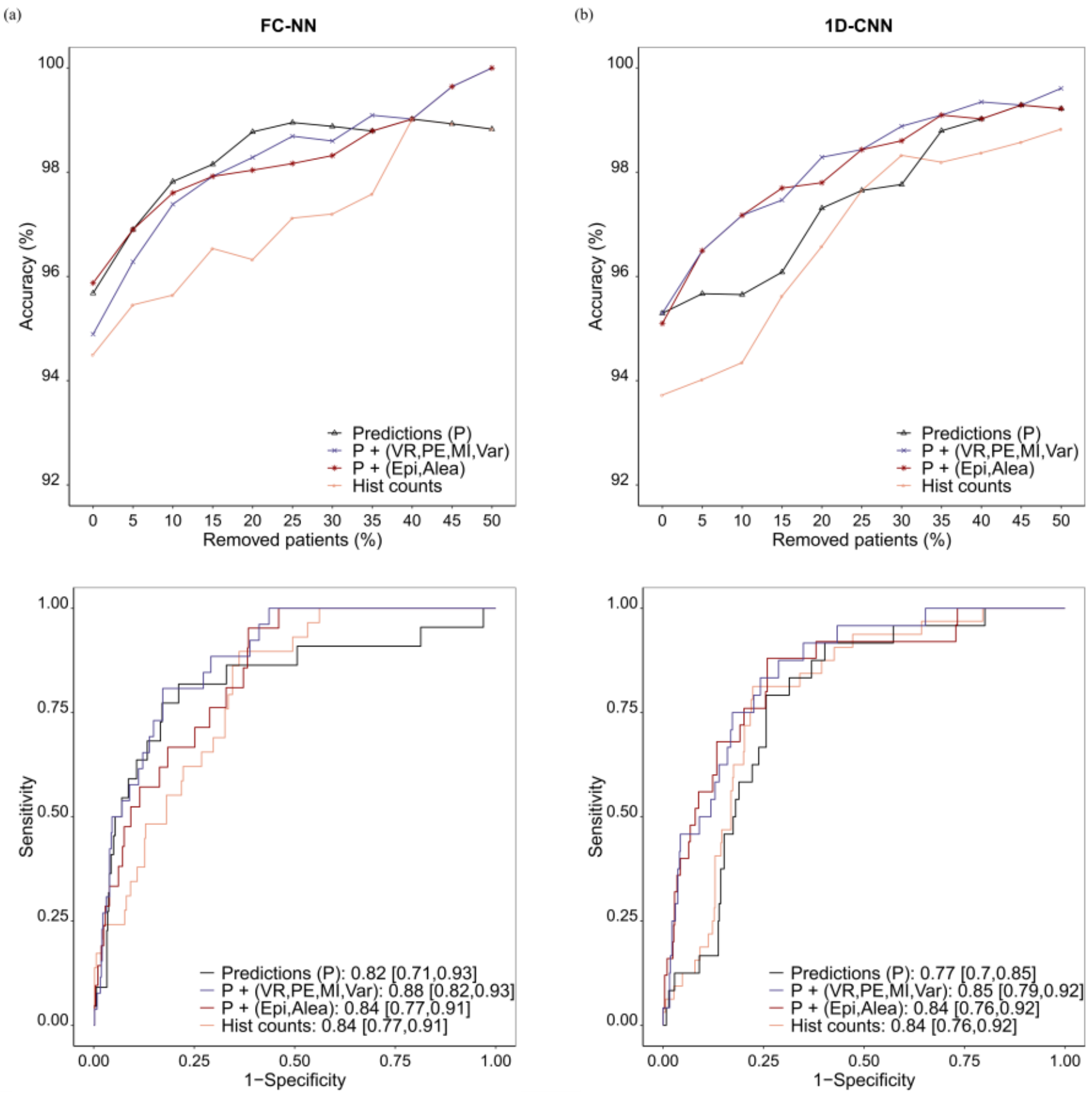

Figure 5. **Uncertainty in terms of the aleatoric uncertainty in patient-level predictions.** The two columns represent the results for the (a) FC-NNs and the (b) 1D-CNNs as aggregation models based on image-level predictions of the 3Ch-MC. The upper row shows the performance improvement when removing uncertain patient-level predictions. The ability of the different aggregation methods to differentiate between correct and erroneous patient-level predictions based on the uncertainty measure is summarized in the lower rows using ROC curves and corresponding AUC values.